\documentclass[a4paper,prd,preprint,tightenlines,floatfix,preprintnumbers,nofootinbib,eqsecnum]{revtex4}

\pdfoutput=1

\usepackage{mathtools}
\usepackage{slashed,color}
\usepackage[absolute,overlay]{textpos} % Allows positioning at specific coordinates

\begin{document}

%\begin{textblock*}{3cm}(0cm,0cm) % width and position (x,y) in cm
\begin{textblock*}{3cm}(17.5cm,.5cm) % width and position (x,y) in cm
    \noindent\includegraphics[width=3cm]{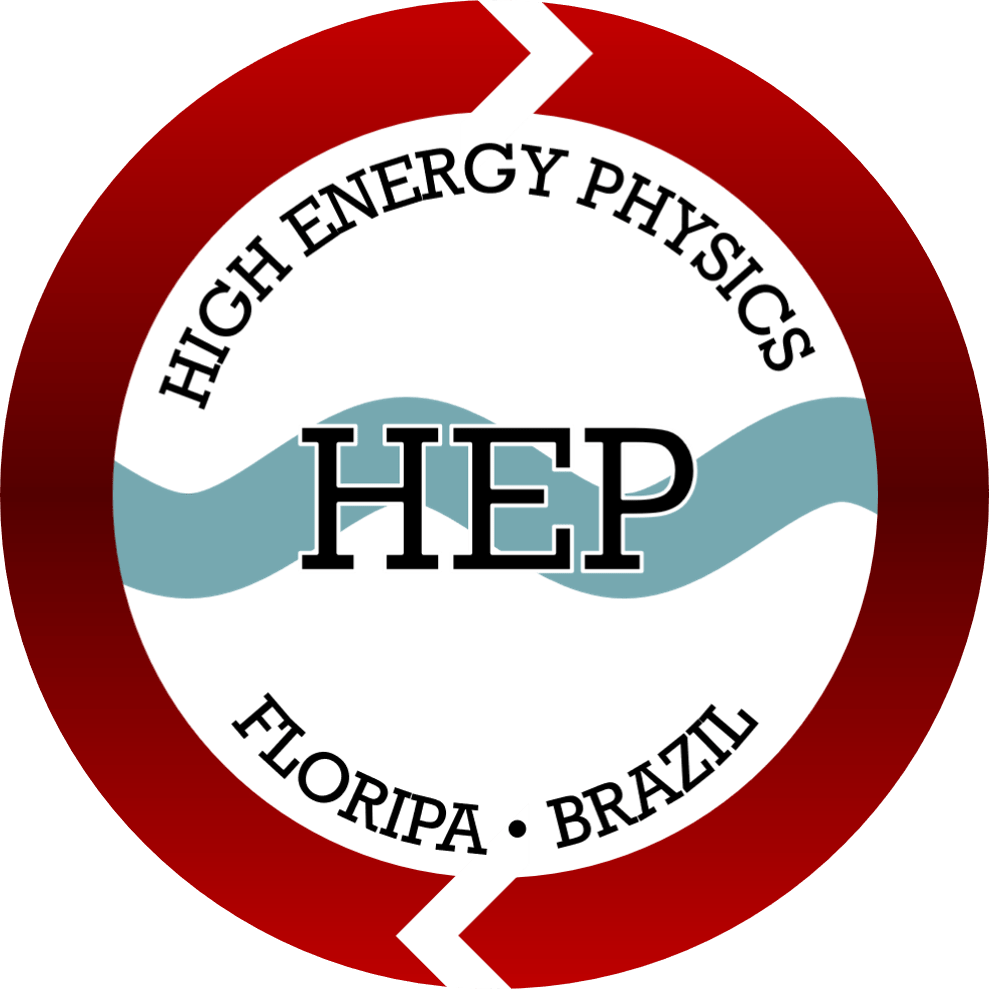} % Replace with your seal image
\end{textblock*}

\title{Predictions for dimuon production in high-energy neutrino--proton collisions using the color dipole model}
\date{\today}

\author{Caetano Ternes$^{1,2}$}
%\email{...}

\author{Daniel Almeida Fagundes$^{3}$}
%\email{...}

\author{Edgar Huayra$^{1}$}
\email{yuberth022@gmail.com}

\author{Emmanuel G. de Oliveira$^{1}$}
\email{emmanuel.de.oliveira@ufsc.br}

\affiliation{
\\
{$^1$\sl Departamento de F\'isica, CFM, Universidade Federal 
de Santa Catarina, C.P. 5064, CEP 88035-972, Florian\'opolis, 
SC, Brazil \\
$^2$\sl Instituto de Física Gleb Wataghin, Universidade Estadual de Campinas, CEP 13083-859, Campinas, SP, Brazil\\
$^3$\sl Department of Exact Sciences and Education, CEE. Federal University of Santa Catarina (UFSC) - Blumenau Campus, 89065-300, Blumenau, SC, Brazil\\
}
}

%----------------------------------------------------------------------
\begin{abstract}
Interactions of high-energy neutrinos with matter can be studied through the angular separation observed in dimuon production, an observable particularly sensitive to the transverse momentum dynamics of partons. In this work, we develop a Monte Carlo event generator based on the color dipole model, interfaced with Pythia8 for parton showering and hadronization simulations, to predict dimuon production cross sections in neutrino--proton collisions at energies relevant to IceCube and future detectors. The color dipole formalism generates larger transverse momentum compared to standard Pythia predictions, enhancing the yield of angularly separated high-energy muons.
\end{abstract}

\maketitle

%----------------------------------------------------------------------
\section{Introduction}

Astrophysical high-energy neutrinos are part of the particle spectrum that reaches Earth. Due to their weak interaction with matter, they can travel through the universe with minimal obstruction. Measuring neutrino-proton cross sections at energies beyond the capabilities of current colliders is a powerful tool for advancing particle physics~\cite{Formaggio:2012cpf, Gaisser:1994yf, Gandhi:1998ri}, providing crucial insights into small-$x$ parton distribution functions~\cite{Candido:2023utz, Xie:2023suk}. It would be valuable to explore observables beyond the inclusive neutrino-proton cross section, allowing us to extract more information from these experiments. For instance, considering the various topologies in the production of taus~\cite{Fagundes:2019wzy} could provide additional insights.

Several neutrino detectors, such as IceCube~\cite{IceCube:2016zyt}, IceCube-Gen2~\cite{IceCube-Gen2:2020qha}, Baikal-GVD~\cite{Baikal-GVD:2023beh}, and KM3NeT 2.0 ARCA~\cite{KM3Net:2016zxf, KM3NeT:2025npi}, detect Cherenkov light radiation emitted by high-energy charged particles (e.g., muons or taus) to observe astrophysical neutrino interactions. These neutrinos have higher energy than the ones measured in ground-based neutrino DIS experiments like NuTeV~\cite{NuTeV:2001dfo}, CCFR~\cite{CCFR:1994ikl}, and NOMAD~\cite{NOMAD:2013hbk}. The Cherenkov light method is particularly effective for detecting high-energy muons, which appear as distinct ``tracks.'' The ability to detect multiple muons in a single event offers the opportunity to explore more exclusive observables~\cite{Barge:2016uzn, Zhou:2021xuh}. To identify multiple muons, a certain angular separation between them is required, considering the detector resolution.

High-energy muons can be produced in the charged-current main vertex, but also in other situations, such as the decay of heavy quarks. Previous studies~\cite{Barge:2016uzn, Zhou:2021xuh} have estimated the cross-section for deep inelastic scattering (DIS) production with angular separation using standard Pythia~\cite{Bierlich:2022pfr} or MadGraph~\cite{Alwall:2014hca} interfaced with Pythia. The main processes that create such dimuons are $\nu_\mu p \rightarrow \mu c \bar{s} X$ and $\nu_\mu p \rightarrow \mu t \bar{b} X$, depending on the neutrino energy. An important background is the W boson production arising from the interaction of neutrinos with nuclear photons, but the DIS contribution is dominant~\cite{Zhou:2021xuh}.

We recognize that the angular separation between muons is an observable highly dependent on the transverse momentum generated at the matrix element (or parton event generator), as well as during showering and hadronization. The larger the angular separation, the larger the transverse momentum. With this in mind, we incorporate, through a new event generator, the physics of the color dipole model~\cite{Kovchegov:2012mbw}, that includes saturation effects and the transverse momentum of the (dipole) quark-antiquark pair. With the help of Pythia8.312 for showering and hadronization, we create a complete Monte Carlo setup with access to all kinematic variables. We do not include nuclear effects at this moment and focus on the proton as a target.

We predict the distributions on the angular separation between muons using our setup. The color dipole event generator gives a nonzero transverse momentum at the matrix element, resulting in dimuons with larger angular separation compared to default leading-order (LO) Pythia8 predictions, as observed in our results. 

The paper is organized into three sections. In Sec.~\ref{sec:DIS}, we provide a brief review of neutrino-proton DIS in the color dipole model and explain how it can be used to build an event generator interfaced with Pythia8. In Sec.~\ref{sec:results}, we discuss the main channels for producing angularly separated muons and present our results. Finally, in Sec.~\ref{sec:summ}, we conclude with a summary of our main findings.

%----------------------------------------------------------------------
\section{Neutrino--proton DIS in the color dipole model}
\label{sec:DIS}

\begin{figure}
    \centering
    \includegraphics[width = 14cm]{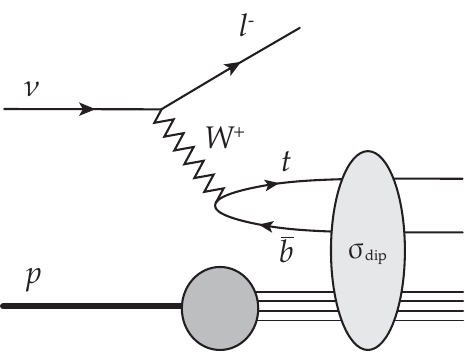}
    \caption{Charged-current deep inelastic scattering $\nu + p \rightarrow l^{-}+ X$ in the color dipole model with $\mu t \bar{b}$ in the final state as an example. Heavy quark ($t,b,c$) decays yield additional muons with significant energy fraction.}
    \label{fig:diagram}
\end{figure}

We will review the high-energy neutrino interaction with a proton through the exchange of a vector boson~\cite{ParticleDataGroup:2024cfk,Devenish:2004pb,Kovchegov:2012mbw}. In deep inelastic scattering (DIS) kinematics, we usually assign the momenta $k$ for the incoming neutrino, $k'$ for the outgoing lepton, $q$ for the exchanged boson and $p$ for the proton. Lepton and proton masses are neglected. The center of mass energy is $s = (p+k)^2$, or in the proton rest frame, $s = 2 m_p E_\nu$, where $m_p$ is the proton mass and $E_\nu$ is the energy of the neutrino. The virtuality of the vector boson is $Q^2$ and $x = Q^2/2 p\cdot q$ is the standard Bjorken variable. The inelasticity parameter is $y = (E_\nu - E_l )/E_\nu$ from $y = Q^2/xs$.

In this context, the neutrino--proton charged-current DIS differential cross section for the production of a heavy quark is given by: 
\begin{equation} \label{eq:dsigmadxdy}
\frac{d^{2} \sigma^{\nu p \rightarrow q\bar{q}'}}{d x d y}
=\frac{s}{4\pi} \frac{G_{F}^{2} M_{W}^{4}}{(M_{W}^{2}+Q^{2})^2}\left[(1-y)^2 F_{+}^{q\bar{q}'} + 2(1-y) F_{0}^{q\bar{q}'} +  F_{-}^{q\bar{q}'}
\right],
\end{equation}
where Fermi constant is $G_F = 1.166 \cdot 10^{-5}$\,GeV$^{-2}$ and the $W$ boson mass is $M_W = 80.4$\,GeV. For an incoming neutrino, a $W^+$ boson is exchanged and the produced pair can be $t \bar{b}$, $c \bar{s}$, etc, as the example shown in Fig.~\ref{fig:diagram}. For an incoming antineutrino, a $W^-$ boson is exchanged, the produced pair can be $b \bar{t}$, $s \bar{c}$, and  $F_{-}$ and $F_{+}$ have to be exchanged in Eq.~\ref{eq:dsigmadxdy}. In the case of the neutral current, the cross section is given by:
\begin{equation} \label{eq:dsigmadxdyNC}
\frac{d^{2} \sigma^{\nu p \rightarrow q\bar{q}}}{d x d y}
=\frac{s}{4\pi} \frac{G_{F}^{2} M_{Z}^{4}}{(M_{Z}^{2}+Q^{2})^2}\left[(1-y)^2 F_{+}^{q\bar{q}} + 2(1-y) F_{0}^{q\bar{q}} +  F_{-}^{q\bar{q}}
\right],
\end{equation}
where the $Z$ boson mass is $M_Z = 91.2$\,GeV. For an incoming antineutrino, $F_{-}$ and $F_{+}$ must again be exchanged.

%For the $W$ boson interaction with the proton, the structure functions with $\lambda = +1, 0, -1$ are proportional to the boson--proton cross sections:
%\begin{align} \label{eq:FW}
%    F^{q\bar{q}'}_\lambda(x, Q^2)
%    = \frac{Q^2}{4 \pi^2} \frac{8}{\alpha_W} \sigma^{Wp}_\lambda 
%    = 2 x {\epsilon_\lambda^\mu}^* {\epsilon_\lambda^\nu} W_{\mu\nu},
%\end{align}
%where ${\epsilon_\lambda^\mu}$ is the boson polarization and $W_{\mu\nu}$ is the hadron tensor. The vertex factor times $g_W^2/8 = 4 \pi \alpha_W/8 = G_F M_W^2 / \sqrt{2}$ was already included in Eq.~\ref{eq:dsigmadxdy}. In the case of a $Z$ boson, an extra factor of 4 appear in the definition:
%\begin{align} \label{eq:FZ}
%    F^{q\bar{q}}_\lambda(x, Q^2)
%    = \frac{Q^2}{4 \pi^2} \frac{2}{\alpha_Z} \sigma^{Zp}_\lambda 
%    = \frac{x}{2} {\epsilon_\lambda^\mu}^* {\epsilon_\lambda^\nu} W_{\mu\nu}.
%\end{align}

For the $W$ boson interaction with the proton, the structure functions with $\lambda = +1, 0, -1$ are proportional to the boson--proton cross sections:
\begin{align} \label{eq:FW}
    F^{W,Z}_\lambda(x, Q^2)
    \propto \frac{Q^2}{4 \pi^2} \sigma^{Wp}_\lambda,
\end{align}
where the proportionality constants depends on what is already included in Eq.~\ref{eq:dsigmadxdy}. In the color dipole model, the structure functions are convolutions over the dipole size $r$ of wave functions and the proton--dipole cross section~\cite{Nikolaev:1990ja,Kutak:2003bd,Fiore:2005yi,Fiore:2011gx}:
\begin{align}
    F_\lambda
    =  \frac{Q^2}{4 \pi^2} \int d^2 r \int_0^1 d z 
    \Psi^\dagger\Psi_\lambda(z, r) \sigma_\text{dip}(\xi , r),
\end{align}
where $z$ is the momentum fraction carried by the quark from the boson. For heavy quarks, the kinematic shift in the momentum fraction is explicitly:
\begin{equation}
\xi = x \left( 1 + \frac{(m_q + m_{\bar{q}'})^2}{Q^2} \right),
\end{equation}
i.e., $\xi/x$ is the minimum of $1 + (m_q^2/z + m_{\bar{q}'}^2/(1-z))/Q^2$.

The $W \rightarrow q \bar{q}'$ wave functions, calculated in the proton rest frame, are given by~\cite{Barone:1992aw}:
\begin{align} 
\label{eq:Wwfrp}
     \Psi^{\dagger} \Psi^{q\bar{q}'}_+(z,r) 
     & = \frac{2 N_c V_{q\bar{q}'}^2}{\pi^2} 
    (1-z)^2 \Big\{ \epsilon^2 K_1(\epsilon r)^2 + m_q^2 K_0(\epsilon r)^2 \Big\} \\
    \label{eq:Wwfrm}
     \Psi^{\dagger} \Psi^{q\bar{q}'}_-(z,r) 
     & = \frac{2 N_c V_{q\bar{q}'}^2}{\pi^2} z^2 
    \Big\{ \epsilon^2 K_1(\epsilon r)^2 + m_{\bar{q}'}^2 K_0(\epsilon r)^2 \Big\}\\
    \label{eq:Wwfr0}
     \Psi^{\dagger} \Psi^{q\bar{q}'}_0(z,r)
     & = \frac{N_c V_{q\bar{q}'}^2}{\pi^2 Q^2} 
    \Big\{( m_q^2 + m_{\bar{q}'}^2) \epsilon^2 K_1(\epsilon r)^2
    + \left( \left( \epsilon^2 + z (1-z) Q^2\right)^2  + m_q^2 m_{\bar{q}'}^2 \right) K_0(\epsilon r)^2 \Big\}
\end{align}
where
\begin{equation}
	\epsilon^2= z(1-z) Q^{2} + (1-z) m_q^2 + z m_{\bar{q}'}^2.
\end{equation}
The general formulas for the wave functions are:
\begin{align} \nonumber
     \Psi^{\dagger} \Psi^{q\bar{q}'}_\pm(z,r) 
     & = \frac{N_c}{2 \pi^2} 
    \Big\{ \left( (g_V \pm g_A)^2 z^2 + (g_V \mp g_A)^2 (1-z)^2 \right) \epsilon^2 K_1(\epsilon r)^2\\     \label{eq:Zwfrp}
    & + \left( g_V ( (1-z) m_q + z m_{\bar q'} ) \mp g_A ( (1-z) m_q - z m_{\bar q'} ) \right)^2 K_0(\epsilon r)^2 \Big\} \\ \nonumber 
     \Psi^{\dagger} \Psi^{q\bar{q}'}_0(z,r)
     & = \frac{N_c }{2\pi^2 Q^2} 
    \Big\{ \left( g_V^2 (m_q - m_{\bar{q}'})^2 + g_A^2 (m_q + m_{\bar{q}'})^2 \right) \epsilon^2 K_1(\epsilon r)^2 \\ \nonumber 
    & + \big( g_V^2 (2 z (1-z) Q^2 + (m_q - m_{\bar{q}'})((1-z)m_q - z m_{\bar{q}'}) )^2 \\     \label{eq:Zwfr0}
    & + g_A^2 (2 z (1-z) Q^2 + (m_q + m_{\bar{q}'})((1-z)m_q + z m_{\bar{q}'}) )^2 \big) K_0(\epsilon r)^2 \Big\}. 
\end{align}
For instance, the $Z \rightarrow q \bar{q}$ wave functions are obtained from  above by setting $q = q'$, $m_q =  m_{\bar{q}'}$, and $g_V$ and $g_A$ to $\frac{1}{2} - \frac{4}{3} \sin^2 \theta_W$ and $- \frac12$ for up-type quarks and to $- \frac12 + \frac23 \sin^2 \theta_W$ and $\frac12$  for down-type quarks.

In this work, since we aim to identify an angular distance between the two produced muons, the integrated cross section is not enough. The vector boson--proton differential cross section is given by (eq. 81 of Ref.~\cite{Kopeliovich:2002yv}):
\begin{align} 
\frac{d \sigma^{Vp}}{d^2 p_t d z} 
= \int d^2 r_1 d^2 r_2 \frac{\mathrm{e}^{i \vec{p}_t \cdot (\vec{r}_1 - \vec{r}_2) }}{(2\pi)^2}
\Psi^\dagger\Psi(z, \vec{r}_1,\vec{r}_2) 
\sigma_\text{dip}(x, z, \vec{r}_1,\vec{r}_2).
\end{align}
In the above formula, we have the quark final transverse momentum $\vec{p}_t$, the quark momentum fraction $z$ and two sizes of dipoles, $\vec{r}_1$ and $\vec{r}_2$. Omitting the $x$ (or $\xi$) dependence, the dipole cross section with two dipole sizes is given by:
\begin{equation}
\sigma_\text{dip}(z, \vec{r}_1, \vec{r}_2)
= \frac{1}{2} \left( 
\sigma_\text{dip}(z \vec{r}_1 + \bar{z} \vec{r}_2) 
+ \sigma_\text{dip}(\bar{z} \vec{r}_1 + z \vec{r}_2) 
- \sigma_\text{dip}(z (\vec{r}_1 - \vec{r}_2)) 
- \sigma_\text{dip}(\bar{z} (\vec{r}_1 - \vec{r}_2)) 
\right).
\end{equation}
The wave functions $\Psi^\dagger\Psi(z, \vec{r}_1,\vec{r}_2)$ can be obtained from Eqs.~\ref{eq:Wwfrp}--\ref{eq:Zwfr0} with the replacement:
\begin{align}
     K_0(\epsilon r)^2 & \rightarrow K_0(\epsilon r_1) K_0(\epsilon r_2)\\
     K_1(\epsilon r)^2 & \rightarrow \cos(\varphi_{12}) K_1(\epsilon r_1) K_1(\epsilon r_2), 
\end{align}
where $\varphi_{12}$ is the angle between $\vec{r}_1$ and $\vec{r}_2$. Actually, the wave functions also have terms proportional to $\sin(\varphi_{12})$ that vanish when integrated over $\vec{r}_1$ and $\vec{r}_2$.

The dipole cross section used is the Golec-Biernat-Wusthoff (GBW) parameterization~\cite{Golec-Biernat:1998zce, Golec-Biernat:1999qor}:
\begin{align}
    \sigma_\text{dip}(x, r) =  \sigma_0 \bigg(1-\exp \left( {-\frac{r^2 Q_s^2(x)}{4}} \right)\bigg),
    \label{eq:GBW}
\end{align}
where $\sigma_0$ is a constant parameter and $Q_s^2(x) = Q_0^2 \big(\frac{x_0}{x} \big)^\lambda$. This parameterization phenomenologically satisfies two key features of the dipole cross section. The first is the well-known color transparency property of dipole scattering ($\sigma_{q \bar{q}} \propto r^2$) in the $r^2 \to 0$ limit. The second is the saturation of the cross section for larger $r$, which becomes relevant in the regime of small $x$, where nonlinear QCD evolution limits the maximum gluon density. Consequently, the saturation scale exhibits a direct dependence on $x$. Ref.~\cite{Mantysaari:2018nng} provides an in depth study of saturation.

The GBW parametrization has the advantage of being a analytical function, making it easy to Fourier transform. The most recent fit of HERA data to the GBW model parameters in Eq.~\eqref{eq:GBW} is presented in Ref.~\cite{Golec-Biernat:2017lfv}:
\begin{align}
    \sigma_0 = 27.43 \, \text{mb}, \, \lambda = 0.248, \, Q_0^2 = 1.0\,\text{GeV}^2, \, x_0 = 4 \times 10^{-5}.
\end{align}
The fit was performed for $x < 0.01$. To achieve the correct behavior at large $x$, we multiplied the dipole cross section by $(1-x)^7$. We use the same quark masses as in the fit, namely, $m_u = m_d = m_s = 0.14$\,GeV, $m_c = 1.4$\,GeV, and $m_b = 4.6$\,GeV. We also use $m_t = 173$\,GeV.

In this work, we use the differential dipole cross section shown above to generate random events with variable weights, effectively constructing a color dipole Monte Carlo event generator. By utilizing the LHAup interface~\cite{Boos:2001cv}, we integrate our code with Pythia Monte Carlo~\cite{Bierlich:2022pfr} version 8.312 for parton showering and hadronization. We use the default settings for Pythia, except the flag ``SpaceShower:dipoleRecoil = on'', as it is the appropriated for DIS. The proper kinematics of the gluon parton, the interacting particle in the dipole approach, are fully incorporated. For the factorization scale, we adopt the smallest transverse mass of the (anti)quark within the dipole.

This setup allows us to access all relevant variables in neutrino–proton collisions, while accounting for the physics of the color dipole model, including saturation effects and the transverse momentum of the quark–antiquark pair. Since our focus is on observables involving muons with angular separation, the generation of transverse momentum—within the matrix element, parton shower, and hadronization—plays a crucial role.

%----------------------------------------------------------------------
\section{Results}
\label{sec:results}

\begin{figure}[thb]
    \centering
    \includegraphics[width=.9\textwidth]{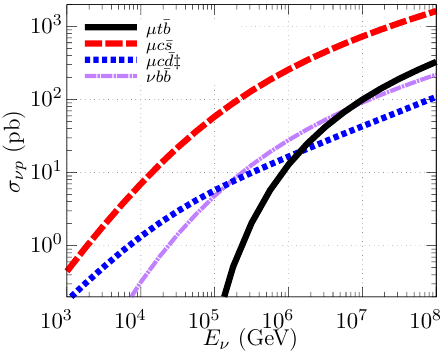}
    \caption{Inclusive cross section for heavy quark production in muon neutrino–proton collisions, shown as a function of neutrino energy and calculated using the color dipole event generator. The main channels for the production of dimuons with angular separation are shown, the labels specify the three particles in the final state of the matrix element. In the case of the $\mu c \bar{d}$ final state, the large valence quark contribution necessitates the use of standard Pythia (without the color dipole model event generator) as indicated by the $\ddagger$ symbol.}
    \label{fig:inclusive}
\end{figure}

\begin{figure}[thb]
    \centering
    \includegraphics[width=.9\textwidth]{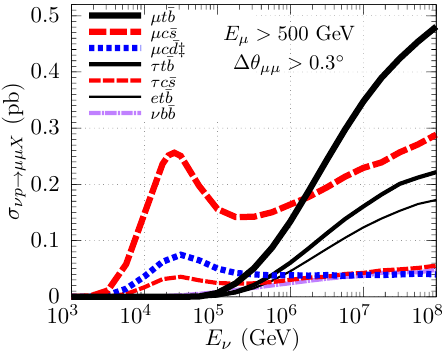}
    \caption{The cross section as a function neutrino energy for the production of pairs of muons with energy larger than 500 GeV and with angular separation larger than 0.3$^\circ$. The main channels are shown, the labels specify the three particles in the final state of the matrix element. The results are calculated using the color dipole event generator, except in the case of the $\mu c \bar{d}$ final state, as the large valence quark contribution necessitates the use of standard Pythia (without the color dipole model event generator) as indicated by the $\ddagger$ symbol.}
    \label{fig:cut-energy}
\end{figure}

\begin{figure}[thb]
    \centering
    \includegraphics[width=.9\textwidth]{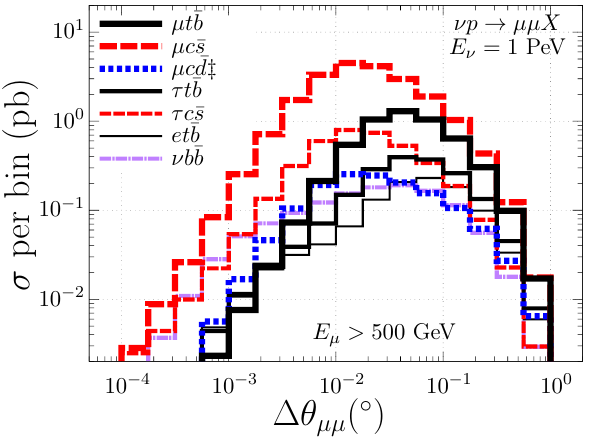}
    \caption{The cross section of a 1 PeV neutrino collision with a proton producing muons with energy greater than 500 GeV calculated using the color dipole event generator is presented as a histogram on the largest angular separation between muons. The labels specify the three particles in the final state of the matrix element. In the case of the $\mu c \bar{d}$ final state, the large contribution from valence quarks necessitates the use of standard Pythia (without the color dipole model event generator) as indicated by the $\ddagger$ symbol.}
    \label{fig:cut-angle}
\end{figure}

\begin{figure}[thb]
    \centering
    \includegraphics[width=.9\textwidth]{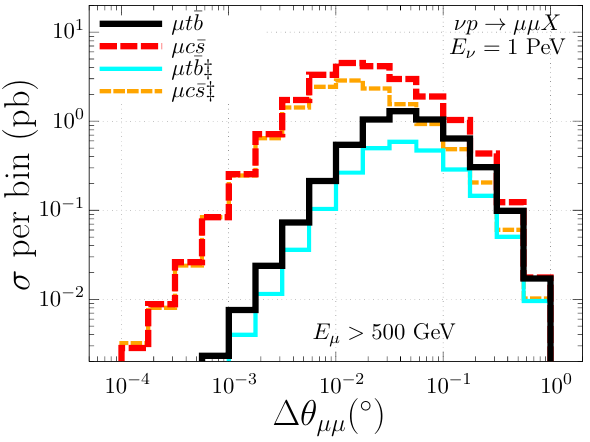}
    \caption{The cross section of a 1 PeV neutrino collision with a proton producing muons with energy greater than 500 GeV is presented as a histogram on the largest angular separation between muons. The labels specify the three particles in the final state of the matrix element. Here, we compare the color dipole model event generator combined with Pythia for showering and hadronization to the standard Pythia ($\mu c \bar{s}\ddagger$) or Pythia plus MadGraph ($\mu t \bar{b}\ddagger$).}
    \label{fig:cut-angle-pythia}
\end{figure}

We are looking for muons with high energy. For example, a muon is produced, via charged current interaction, when the incoming particle is a muon neutrino. This production occurs at the matrix element level, with the resulting muon typically carrying a significant fraction of the neutrino energy. In the case of an incoming tau neutrino, a tau lepton is produced, which can subsequently decay into a muon, also carrying a substantial portion of the initial energy.

On the other hand, the production of heavy quarks ($c, b, t$) often leads to the creation of muons in the final state of the collision, as these quarks are primarily produced via the strong force and decay via the weak force. Therefore, processes involving an incoming tau or even electron neutrino can result in the production of high-energy muons if heavy quarks are generated. Additionally, neutral current interactions can produce muons through the production of heavy quark pairs. The presence of heavy quarks also ensures that the color dipole approach works effectively, as it is not intended for large-$x$ light valence quarks.

We focus on identifying high-energy muons by selecting those with $E_{\mu} > 500$\,GeV. This threshold is chosen because the energy distribution of detected muons in IceCube is significantly suppressed by selection cuts for muon energies below approximately 500\,GeV~\cite{IceCube:2021xar}. Additionally, we require that two high-energy muons have an angular separation greater than 0.3$^\circ$. This criterion is based on the angular resolution considerations in neutrino telescopes, which are influenced by the spacing between the digital optical modules responsible for detecting Cherenkov photons~\cite{Zhou:2021xuh, Groom:2001kq}. These cuts make our observable highly sensitive to the transverse momentum of the muons.

We present in Fig.~\ref{fig:inclusive} the inclusive cross section for heavy quark production in muon neutrino–proton collisions, as a function of neutrino energy. Only the main channels for the dimuon observable described above are shown, labeled by the three particles in the final state. Cases with incoming electron or tau neutrinos (and with an electron or tau instead of a muon in the final state) have nearly the same inclusive cross section as the incoming muon neutrino. Other cases involving heavy quarks are not shown, as they are suppressed by the small CKM matrix elements: $|V_{td}|^2, |V_{ts}|^2, |V_{ub}|^2, |V_{cb}|^2 \ll 1$. There is one exception: the case of the $\mu c \bar{d}$ final state, the large valence $d$ quark contribution necessitates the use of Pythia without color dipoles, as indicated by the $\ddagger$ symbol.

Cases involving only light quarks ($u \bar{d}$ or $u \bar{s}$) could produce a large inclusive cross section, but they typically do not yield more than one high-energy muon and are therefore excluded. Neutral current interactions generally have smaller cross sections, and we include only the most relevant case, $\nu b \bar{b}$. The $\nu c \bar{c}$ case produces fewer muons, as the $c$ quark has a shorter decay chain than the $b$ quark, and the $\nu t \bar{t}$ case is suppressed by the large $t$ quark mass at the energies shown.

In Fig.~\ref{fig:cut-energy}, the cross section as a function of neutrino energy for the production of the dimouns with the cuts explained above is shown. We see that the \(W^+ \rightarrow c\bar{s}\) and \(W^+ \rightarrow t\bar{b}\) channels  are the most relevant, at $E_\nu = 10^4 \sim 10^6$\,GeV and $E_\nu = 10^6 \sim 10^8$\,GeV, respectively.  All the other channels that give sizable contributions are also shown.

In order to better understand our predictions, we present in Fig.~\ref{fig:cut-angle} the cross-section histogram for the largest angular separation between muons with energy greater than 500 GeV. We observe that the cases with a $t\bar{b}$ final state are skewed toward larger angular separations compared to the $c\bar{s}$ cases. This occurs because the more massive the quark, the larger the transverse momentum imparted to the muon from the quark decay.

In Fig.~\ref{fig:cut-angle-pythia}, we compare the color dipole model event generator combined with Pythia for showering and hadronization to the standard Pythia ($\mu c \bar{s}\ddagger$) or Pythia plus MadGraph ($\mu t \bar{b}\ddagger$). As in Fig.\ref{fig:cut-angle}, we present the cross-section histogram for a 1 PeV neutrino. The cross section from the color dipole model is slightly larger and, in the case of the $\mu c \bar{s}$ final state, visibly shifted toward larger angular separations. This shift arises because the dipole model incorporates $p_t$ dependence in the dipole wave function, resulting in additional transverse momentum for the final heavy quarks.

%----------------------------------------------------------------------
\section{Summary}
\label{sec:summ}

We have predicted the cross sections for high-energy dimuon production in interactions between high-energy neutrinos and protons, using the color dipole model in conjunction with Pythia8 Monte Carlo showering and hadronization simulations. The primary source of high-energy muons, aside from the incoming neutrino, is the heavy quarks produced via QCD interactions. The main channels identified are \(W^+ \rightarrow c\bar{s}\) and \(W^+ \rightarrow t\bar{b}\), at lower and higher energies respectively.

Our predictions include the dependence on the angle between the muons, as this observable is particularly sensitive to transverse momentum generated in the hard matrix element or parton evolution, with a minimum angular separation required for individual muon identification. The color dipole formalism predicts a nonzero transverse momentum at the matrix element, resulting in dimouns with larger angular separation compared to default LO Pythia8 predictions, as observed in the \(c\bar{s}\) case. 

Looking ahead, NLO Monte Carlo predictions~\cite{Buonocore:2023kna,vanBeekveld:2024ziz} would be valuable for comparison with our results. Additionally, including nuclear effects such as shadowing would enhance the current analysis. Direct dimuon detection by IceCube and future detectors will further validate our predictions and help constrain Monte Carlo event generators.

%----------------------------------------------------------------------
\section*{Acknowledgments}

This work was supported by Fapesc, INCT-FNA (464898/2014-5), and CNPq (Brazil) for CT, DAF, EH, and EGdO. This study was financed in part by the Coordenação de Aperfeiçoamento de Pessoal de Nível Superior -- Brasil (CAPES) -- Finance Code 001. 

%----------------------------------------------------------------------

%----------------------------------------------------------------------
\end{document}